\documentclass[]{revtex4}
\usepackage{hyperref}
\usepackage{amsmath}
\usepackage[psamsfonts]{amssymb }
\usepackage{mathrsfs}

\begin{document}

\title{Finite field-dependent symmetries in perturbative quantum gravity}

\author{ Sudhaker Upadhyay}
 \email {  sudhakerupadhyay@gmail.com; 
 sudhaker@boson.bose.res.in}

\affiliation { S. N. Bose National Centre for Basic Sciences,\\
Block JD, Sector III, Salt Lake, Kolkata -700098, India. }
 
\begin{abstract}
In this paper we discuss the 
 absolutely anticommuting nilpotent symmetries for  perturbative quantum gravity in general curved spacetime in linear and non-linear gauges. Further, we  
 analyze the finite field-dependent BRST (FFBRST) transformation 
for   perturbative quantum gravity in general curved spacetime. The FFBRST transformation changes the gauge-fixing and ghost parts of the
perturbative quantum gravity  within functional integration. However, the operation of
such symmetry transformation  on the generating functional  of perturbative quantum gravity  does  not  
affect the theory on physical ground.
The FFBRST transformation with appropriate choices of finite BRST parameter connects non-linear Curci--Ferrari and Landau gauges
 of perturbative quantum gravity. 
The validity of the results is also established at quantum level using Batalin-Vilkovisky (BV) formulation.
 
  \end{abstract}
\maketitle

\section{  Introduction}
The study of the structure of spacetime at Planck scale, where  the quantum   effects
of gravity cannot be neglected, is a great challenge for fundamental physics.	It is  very essential
to understand the perturbative quantum gravity  for those
who want  to proceed towards any kind of non-perturbative approach \cite{hooft}.
 The  perturbative quantum gravity in curved spacetime  as a gauge theory is a   subject of 
 interest form many respects \cite{hata1, hata2, asch}.
  The mode analysis and Ward identities for a ghost propagator for perturbative quantum gravity in de Sitter space 
  has been discussed iteratively  \cite {tsa, fi}.
The Feynman rules and propagator for gravity in the presence of a flat Robertson--Walker background  in the physically interesting cases of inflation have been analyzed \cite{wood}.
 Such models of gravity have founded great attempts to unify gravity with Maxwell theory \cite{ein}. The gravity models
 with gauge invariance have their relevance in string theories also \cite{ch,da,ah}.

The quantum theory of gravity in general curved spacetime has general coordinate (gauge) invariance and hence cannot
be quantized without getting rid of the redundant degrees of freedom. This can be achieved 
by imposing a suitable gauge conditions. The Landau   and non-linear Curci--Ferrari gauge conditions
play a pivotal role in the analysis of gauge and ghost condensation of the perturbation theory
\cite{dud1, dud2}.
 These gauge  conditions can be incorporated to the theory of gravity
at quantum level by adding the suitable  gauge-fixing and ghost terms to the classical action which remains 
invariant under the fermionic rigid
 BRST  invariance \cite{brst, tyu}.
 The BRST symmetry plays an  important role to study the unitarity and renormalizability of the
gauge theories \cite{ht,wei}.  
However, BV formulation  to quantize the more general 
gauge theories with open gauge algebra is more fundamental approach to study the supergravity   and topological field theories \cite{ht,wei,bv,bv1,bv2,subm, bss}.
The BRST and the anti-BRST symmetries for perturbative quantum gravity
in  flat spacetime dimensions have been studied by many people
\cite{na,ku,ni} and their work has been summarized by N. Nakanishi and I. Ojima \cite{nn}.
The BRST symmetry in two dimensional curved spacetime has been thoroughly
studied \cite{yo,be,fr}.  Recently, the BRST formulation in the theory of perturbative quantum gravity has been analyzed  \cite{faiza, upa}. 
 The BRST  symmetry transformations of the gauge theories in flat spacetime have been generalized by making the parameter finite  and field-dependent  which is known as FFBRST transformations \cite{sdj}. The FFBRST transformations have found several applications in gauge field theories 
 in flat spacetime \cite{sdj,sdj1,rb,susk,sb,smm,fs,sud1,sudhak,rbs,rs}.
 However, so far FFBRST formulation has not been developed for any
 theory of curved spacetime. This
 provides a motivation to develop
 FFBRST transformation in curved spacetime.
 We develop such  a
formulation for the first time for the theory of quantum gravity in the curved spacetime. 

In this paper we discuss the BRST and anti-BRST invariance of
gravity theory in Landau and massless Curci--Ferrari gauges. Further, we
investigate the FFBRST transformation for perturbative quantum gravity.
The FFBRST transformation is constructed by replacing the
infinitesimal field-independent BRST parameter with a finite field-dependent global parameter.
The formal aspects of such FFBRST formulation are discussed with full generality,
in which we show that the FFBRST transformation is symmetry of the action, however,
it does not leave the path integral measure of functional integral invariant.
The explicit form of the non-trivial Jacobian of the  path  integral measure is calculated for 
the theory of quantum gravity. 
The non-trivial Jacobian  changes the gauge-fixing 
and ghost terms within the functional integral of perturbative quantum gravity. 
We explicitly show that for a proper choice of field-dependent  parameter
the FFBRST transformation connects the linear and non-linear gauges within the functional integration 
of perturbative quantum gravity. 
The results are also tested at quantum level using BV formulation 
of perturbative quantum gravity.

This paper is organized as follows. In Sec. II, we discuss the different gauge conditions in
perturbative gravity with BRST invariance. In Sec. III, we develop the field-dependent BRST 
symmetry in the theory of curved spacetime and show that
Landau and non-linear gauges can be connected with suitable choices of finite parameter.
The result is also established at quantum level in section IV.
In the last section, we summarize the results with some discussion on future investigations.

\section{ The perturbative quantum gravity }
We start with the classical Lagrangian density for  gravity in general curved spacetime  
\begin{equation} 
{\cal L}_c   =  \frac{\sqrt{ -g}}{16\pi G}(R-2\lambda), \label{kin}
\end{equation}
where $R$ is Ricci scalar curvature and   $\lambda$ is  a cosmological constant.
 
 In this theory the full metric $g_{ab}^f$ is written in terms of a fixed
  metric of background spacetime  $g_{ab}$ and small perturbations around it. 
The small perturbation around the fixed background metric, denoted by  $h_{ab}$, is 
considered as a  field that is to be quantized. 
So, we can write 
\begin{equation}
g_{ab}^f=g_{ab}+h_{ab}.
\end{equation}
 The Lagrangian density given in Eq. (\ref{kin}) remains invariant
 under the following general coordinate transformation, which is infinitesimal in nature,
 \begin{eqnarray}
 \delta_\Lambda h_{ab}=\nabla_a \Lambda_b +\nabla_b \Lambda_a + {\pounds}_{(\Lambda)} h_{ab}.
 \end{eqnarray}
The Lie derivative of  $h_{ab}$ with respect to the vector field $\Lambda_a$ is given  by
 \begin{eqnarray}
 {\pounds}_{(\Lambda)} h_{ab}=\Lambda^c\nabla_c h_{ab} +h_{ac}\nabla_b \Lambda^c
 +h_{ cb}\nabla_a \Lambda^c.
 \end{eqnarray}
As the theory for perturbative quantum gravity is gauge invariant it contains some redundant degrees 
 of freedom.  These redundancy of degrees of 
freedom give rise to constraints in the canonical quantization  \cite{ht} and 
divergences in the generating functional   in the path integral  quantization. In 
order to remove the 
redundancy in degrees of freedom we restrict the gauge by following  
gauge-fixing condition  \begin{equation}
G[h]_a=(\nabla^b h_{ab} -k\nabla_a h) =0,
\end{equation}
where $k\neq  1$. For $k=1$ the conjugate momentum corresponding to $h_{00}$ 
vanishes
and therefore the partition function diverges again. 
For this reason sometimes $k$ is written in terms of an arbitrary finite constant $\beta$ as $1+\beta^ {-1}$ \cite{hig}.
To ensure the unitarity of the perturbative quantum gravity a 
Faddeev--Popov  ghost term is also needed.
 
The effects of above gauge condition can be incorporated in the theory
by adding suitable gauge-fixing   and corresponding ghost terms  
in the classical Lagrangian density given in Eq. (\ref{kin}).
For this theory the Landau gauge-fixing and corresponding Faddeev--Popov ghost terms 
have the following form:
  \begin{eqnarray}
{\cal L}_{gf}&=&  \sqrt{- g}[ib^a(\nabla^b h_{ab}-k \nabla_a h)], \label{gfix}\\
{\cal L}_{gh}&=& i \sqrt{- g}\bar c^a \nabla^b [ \nabla_a c_b+ \nabla_b c_a- 
2kg_{ab}\nabla_c c^c +
  {\pounds}_{(c)} h_{ab}  -kg_{ab}g^{cd} {\pounds}_{(c)} h_{cd}], \nonumber\\
  &=&\sqrt{ -g}\bar c^a M_{ab} c^b,
\end{eqnarray} 
with Faddeev--Popov matrix operator $M_{ab}$,  explicitly,  defined as
\begin{equation} 
M_{ab}= i  \nabla_c [ \delta_b^c\nabla_a  + g_{ab}\nabla^c - 2k \delta_a^c\nabla_b 
+\nabla_b h^c _a -h_{ab}\nabla^c
-h^c_b\nabla_a 
-kg^c_ag^{ef}(\nabla_b h_{ef} +h_{eb}\nabla_f +h_{fb}\nabla_e)].
 \end{equation}
Here we note that in the theory of perturbative gravity the Faddeev-Popov ghost and anti-ghost fields  are vector fields.

Now, the complete effective action for perturbative quantum gravity in four curved
spacetime dimensions (in Landau gauge) is written as
\begin{equation}
S_L =\int d^4 x ({\cal L}_c +{\cal L}_{gf}+{\cal L}_{gh}),  \label{com}
\end{equation}
which remains invariant under following BRST variations of fields,  
\begin{eqnarray}
s  h_{ab} =   (\nabla_a c_b +\nabla_b c_a +{\pounds}_{(c)} h_{ab}), \  
s c^a  =  -c_b\nabla^b c^a,     \ s  \bar c^a
= b^a,\ s  b^a  =  0.\label{sym}
\end{eqnarray}
This  effective action is also invariant under the following anti-BRST
transformations where the roles of ghost and anti-ghost fields are interchanged, 
\begin{eqnarray}
\bar s  h_{ab} =   (\nabla_a \bar c_b +\nabla_b \bar c_a +{\pounds}_{(\bar c)} 
h_{ab}), \  
\bar s \bar c^a  =  -\bar c_b\nabla^b \bar c^a,     \ \bar s    c^a
= -b^a,\ \bar s  b^a  =  0.\label{ansym}
\end{eqnarray}
The above BRST and anti-BRST transformations are nilpotent in nature and satisfy 
absolute  anticommutivity, i.e. 
\begin{eqnarray}
s^2=0,\ \bar s^2 =0,\ \{s, \bar s \} =0.
\end{eqnarray} 
 
Now, we express the gauge-fixing and ghost part of the 
complete Lagrangian density as follows,
\begin{eqnarray}
{\cal L}_g &=& {\cal L}_{gf} +{\cal L}_{gh},\nonumber\\
&=&i s \sqrt{ -g}[\bar c ^a (\nabla^b h_{ab} -k\nabla_a h)],\nonumber\\
&=&-i \bar s \sqrt{- g}[ c ^a (\nabla^b h_{ab} -k\nabla_a h)],\nonumber\\
&=&-\frac{1}{2}i s\bar s \sqrt{- g}  ( h^{ab} h_{ab}  ),\nonumber\\
&=&\frac{1}{2}i \bar s s\sqrt{- g}  ( h^{ab} h_{ab}  ).\label{g}
\end{eqnarray}
In the BV formalism, the gauge-fixing and ghost part of the Lagrangian density   
is generally expressed in terms of BRST variation of a gauge-fixed fermion.
It is straightforward to write the ${\cal L}_{g }$ given in Eq. (\ref{g})
in terms of gauge-fixed fermion $\Psi$ as    
\begin{equation}
{\cal L}_{g }=s  \Psi,
\end{equation}
where the expression for $\Psi$ is 
\begin{equation}
\Psi =i  \sqrt{- g}[\bar c ^a (\nabla^b h_{ab} -k\nabla_a h)].\label{gff}
\end{equation}

In non-linear Curci--Ferrari gauge condition the 
gauge-fixing and ghost terms can be expressed as 
\begin{eqnarray}
\mathcal{L'}_g&=& {\cal L'}_{gf} + {\cal L'}_{gh},\nonumber\\
&=&  \sqrt{ -g}\left[ib^a(\nabla^b h_{ab}-k \nabla_a h)-i\bar c^b \nabla_b c^a (\nabla^b h_{ab}-k \nabla_a h)+\bar c^a M_{ab} c^b +\frac{\alpha}{2}b^b\nabla_b \bar c^a c_a\right.\nonumber\\
&-&\left. \frac{\alpha}{2} \bar c^c\nabla_c c^b \nabla_b \bar c^a c_a -\frac{\alpha}{2} \bar b^b \nabla_b b^a c_a -\frac{\alpha}{2}\bar c^b\nabla_b \bar c^a c_d\nabla^d c_a -\frac{\alpha}{2} b_a b^a +\alpha \bar c^b b^b \nabla_b c_a\right.\nonumber\\
&+&\left. \alpha \bar c^a \bar c^b c^d \nabla_b \nabla_d c_a\right],\label{nlg}
\end{eqnarray}
where $\alpha$ is a gauge parameter. 
With these gauge-fixing and Faddeev--Popov ghost terms  the effective action of perturbative quantum gravity in non-linear gauge is written as
\begin{equation}
S_{NL} =\int d^4 x ({\cal L}_c +{\cal L'}_{g}).  
\end{equation}
The BRST transformations  for perturbative quantum gravity  in Curci--Ferrari gauge are given by
\begin{eqnarray}
s \,h_{ab} &=& \nabla_a c_b + \nabla_b c_a + \pounds_{(c)} h_{ab}, \nonumber \\
s \,c^a &=& - c_b \nabla^b c^a, \nonumber \\
s \,\bar{c}^a &=& b^a - \bar{c}^b\nabla_b c^a, \nonumber \\ 
s \,b^a &=& - b^b\nabla_b c^a -  \bar{c}^b c^d\nabla_b \nabla_d c^a,\label{nlbrs}
\end{eqnarray}
and the anti-BRST symmetry transformations for this theory are constructed as
\begin{eqnarray}
\bar{s}\, h_{ab} &=& \nabla_a \bar{c}_b + \nabla_b \bar{c}_a + \pounds_{(\bar{c})} h_{ab}, \nonumber \\
\bar{s} \,\bar{c}^a &=& - \bar{c}_b \nabla^b \bar{c}^a, \nonumber \\
\bar{s} \,c^a &=& - b^a - \bar{c}^b\nabla_b c^a, \nonumber \\ 
\bar{s} \,b^a &=& - b^b\nabla_b \bar{c}^a +  c^b\bar{c}^d\nabla_b\nabla_d  \bar{c}^a.
\end{eqnarray}
These BRST and anti-BRST transformations are also absolutely anticommuting and nilpotent in nature. 
Now, we  are able to write the non-linear  gauge-fixing and ghost part of the effective Lagrangian 
density given in Eq. (\ref{nlg}) as  
\begin{eqnarray}
\mathcal{L'}_g&=&i s \sqrt{- g}\left[\bar c ^a \left(\nabla^b h_{ab} -k\nabla_a h-i\frac{\alpha}{2}\nabla_a \bar c^b c_b 
+i\frac{\alpha}{2} b_a +i\frac{\alpha}{2}\bar c^b \nabla_b c_a\right)\right],\nonumber\\
&=&- \frac{i}{2}s\bar{s}\sqrt{-g}\left[h^{ab}h_{ab} - i \alpha \bar{c}^a c_a \right], \nonumber \\ &=&\frac{i}{2}\bar{s} s\sqrt{-g} \left[h^{ab}h_{ab} - i \alpha \bar{c}^a c_a \right].
\end{eqnarray}
We will study the generalization of such nilpotent symmetries in the next section. 
\section{FFBRST formulation for perturbative quantum gravity}
To develop the FFBRST formulation for theory of quantum gravity in curved spacetime we start with the
usual BRST transformation written in terms of infinitesimal and field-independent
Grassmann parameter  $\delta\Lambda$ as
\begin{equation}
\delta_b \phi (x) =s\phi (x) \delta \Lambda,
\end{equation} 
where $\phi (x)$ is the generic notation of fields $(h, c, \bar c, b)$ involved the theory of quantum gravity.
The properties of the such BRST transformation  do not depend on whether 
the parameter $\delta\Lambda$  is (i) finite or infinitesimal, (ii) field-dependent or not, as long 
as it is anticommuting and spacetime independent. These observations give us a liberty to 
generalize the BRST transformation by making the parameter, $\delta\Lambda$ finite and field-dependent without
 affecting its properties.  To do so, we 
start by interpolating   a continuous parameter, $\kappa\ (0\leq \kappa\leq 1)$, in the theory
 to make the  infinitesimal parameter field-dependent.
We allow the fields, $\phi(x,\kappa)$, to depend on  $\kappa$  in such a way that $\phi(x,\kappa =0)=\phi(x)$ the initial fields and $\phi(x,\kappa 
=1)=\phi^\prime(x)$, the transformed fields.

We consider the intermediate fields $\phi(x,\kappa), (0\leq \kappa\leq 1)$ satisfying following infinitesimal
field-dependent BRST transformation  \cite{sdj}
\begin{equation}
{d\phi(x,\kappa)}=s  [\phi (x) ]\Theta^\prime [\phi ( \kappa ) ]{d\kappa},
\label{diff}
\end{equation}
where the $\Theta^\prime [\phi ( \kappa ) ]{d\kappa}$ is the infinitesimal but field-dependent parameter.
The FFBRST transformation with the finite field-dependent parameter then can be 
constructed by integrating such infinitesimal transformation from $\kappa =0$ to $\kappa= 1$, to obtain
\begin{equation}
\phi^\prime\equiv \phi (x,\kappa =1)=\phi(x,\kappa=0)+s [\phi(x) ]\Theta[\phi  ],
\label{kdep}
\end{equation}
where 
\begin{equation}
\Theta[\phi ]=\int_0^1 d\kappa^\prime\Theta^\prime [\phi( \kappa^\prime)],
\end{equation}
 is the finite field-dependent parameter. 
Such   transformations with finite field-dependent
 parameter are the symmetry  of the effective action but not of the 
 functional integral \cite{sdj} as the 
path integral measure  is not invariant under such transformations. Thus the Jacobian 
of path integral measure gives some non-trivial contribution to the generating functional of the theory.

The Jacobian of the path integral measure for such transformations is then evaluated for some 
particular choices of the finite field-dependent parameter, $\Theta[\phi(x)]$, as
\begin{eqnarray}
{\cal D}\phi &=&J(
\kappa) {\cal D}\phi(\kappa).
\end{eqnarray}
We substitute the Jacobian, $J(\kappa )$,   within the functional integral  as
\begin{equation}
J(\kappa )\rightarrow \exp[iS_1[\phi(x,\kappa), \kappa ]],\label{s}
\end{equation}
where $ S_1[\phi (x), \kappa]$ is local functional of fields.
This imposes  the following condition to the theory to be satisfied  \cite{sdj}
 \begin{eqnarray}
  \langle\langle \frac{1}{J}\frac{dJ}{d\kappa}-i\frac
{dS_1[\phi (x,\kappa ), \kappa]}{d\kappa} \rangle \rangle_\kappa  =0. \label{mcond}
\end{eqnarray}

In this method we calculate the infinitesimal change in the $J(\kappa)$
with the help of following  condition
\begin{equation}
\frac{1}{J}\frac{dJ}{d\kappa}=-\int d^4y\left [\pm s\phi (y,\kappa )\frac{
\delta\Theta^\prime [\phi ]}{\delta\phi (y,\kappa )}\right],\label{jac}
\end{equation}
where sign $+$ is used for bosonic fields $\phi$ and   $-$ sign is used for  fermionic
fields $\phi$.

\subsection{From non-linear gauge to Landau gauge}
The FFBRST transformation   for perturbative quantum gravity  in  the massless Curci--Ferrari gauge  is 
constructed as
\begin{eqnarray}
f \,h_{ab} &=& (\nabla_a c_b + \nabla_b c_a + \pounds_{(c)} h_{ab})\ \Theta[\phi], \nonumber \\
f \,c^a &=& - c_b \nabla^b c^a\ \Theta[\phi], \nonumber \\
f \,\bar{c}^a &=& (b^a - \bar{c}^b\nabla_b c^a) \ \Theta[\phi], \nonumber \\ 
f \,b^a &=& (- b^b\nabla_b c^a -  \bar{c}^b c^d\nabla_b \nabla_d c^a)\ \Theta[\phi],\label{ff}
\end{eqnarray}
where $\ \Theta[\phi]$ is finite field-dependent parameter. To 
connect the Landau and non-linear Curci--Ferrari gauge we construct the finite parameter
obtainable from following infinitesimal field-dependent parameter
\begin{equation}
\Theta'[\phi] =-i\frac{\alpha}{2}\sqrt{-g}\int d^4 y\ (\bar c_b \nabla^b\bar c ^a c_a -\bar c^a b_a -\bar c^a 
\bar c_b \nabla^b c_a).\label{theta}
\end{equation}
For this expression of $\Theta'$ and BRST given in Eq. (\ref{nlbrs}) the change in Jacobian,  using Eq. (\ref{jac}), 
is calculated as follows 
\begin{eqnarray}
\frac{1}{J(\kappa)}\frac{dJ(\kappa)}{d\kappa}&=& i\frac{\alpha}{2}\sqrt{-g}\int d^4x \left[
-b_b\nabla^b \bar c^a c_a + \bar c^d \nabla_d c_b\nabla^b\bar c^a c_a
+\bar c_b \nabla^b b^a c_a +\bar c_b \nabla^b \bar c^a c_d\nabla^d c_a\right.
\nonumber\\
&+&\left. b_a b^a -2\bar c^a b_b\nabla^b c_a
-2\bar c^a \bar c_b c_d \nabla^b\nabla^d c_a\right].\label{j}
\end{eqnarray}
The local functional $S_1$ appearing in the Eq. (\ref{s}) is written to have the following explicit form
\begin{eqnarray}
S_1[\phi(\kappa), \kappa]&=& \int d^4x \left[\xi_1 
 b_b\nabla^b \bar c^a c_a +\xi_2 \bar c^d \nabla_d c_b\nabla^b\bar c^a c_a
+\xi_3\bar c_b \nabla^b b^a c_a +\xi_4\bar c_b \nabla^b \bar c^a c_d\nabla^d c_a\right.
\nonumber\\
&+&\left.\xi_5 b_a b^a +\xi_6 \bar c^a b_b\nabla^b c_a
+\xi_7 \bar c^a \bar c_b c_d \nabla^b\nabla^d c_a\right],
\end{eqnarray}
where all fields and $\xi_i (i=1,2,..,7)$, involved in the above expression, depend on parameter $\kappa$.
Now we have to identify the exact values of $\xi_i$ in terms of $\kappa$. For this purpose, we
 calculate the change in $S_1$ with the help of Eq. (\ref{diff}) as 
 \begin{eqnarray}
\frac{dS_1[\phi(\kappa), \kappa]}{d\kappa}&=& \int d^4x \left[\xi_1' 
 b_b\nabla^b \bar c^a c_a +\xi_2' \bar c^d \nabla_d c_b\nabla^b\bar c^a c_a
+\xi_3'\bar c_b \nabla^b b^a c_a +\xi_4'\bar c_b \nabla^b \bar c^a c_d\nabla^d c_a\right.
\nonumber\\
&+&\left.\xi_5' b_a b^a +\xi_6' \bar c^a b_b\nabla^b c_a
+\xi_7' \bar c^a \bar c_b c_d \nabla^b\nabla^d c_a -(\xi_1+\xi_2)(b^c\nabla_c c_b\nabla^b\bar c^ac_a \right.\nonumber\\
&+&\left. \bar c^c c^d\nabla_c\nabla_d c_b\nabla^b \bar c^a c_a )\Theta' -(\xi_1 +\xi_3 ) b_b \nabla^b b^a c_a\Theta' -(\xi_1 +\xi_4) b_b \nabla^b \bar c ^a c_d \nabla^d c_a \Theta' \right.
\nonumber\\
&-&\left. (\xi_2 -\xi_3)\bar c^d\nabla_d c_b\nabla^b b^a c_a\Theta' -(\xi_2 -\xi_4) \bar c^d \nabla_d c_b \nabla^b \bar c^a c_c\nabla^c c_a \right.
\nonumber\\
&-&\left. (\xi_3-\xi_4) \bar c_b \nabla^b b^a c_c\nabla^c c_a \Theta' -(2\xi_5 +\xi_6 )b^a b^b\nabla_b c_a\Theta'
-(2\xi_5 +\xi_7) b^a\bar c^b c^d\nabla_b\nabla_d c_a \Theta' \right.\nonumber\\
&+&\left.(\xi_6 -\xi_7 )(\bar c^a\bar c^c c^d \nabla_c\nabla_d c^b \nabla_b c^a -\bar c^a b^b c^d \nabla_b \nabla_d c_a )\Theta'
\right],\label{dis}
\end{eqnarray}
where prime denotes the derivative with respect to $\kappa$. Now, the condition given in (\ref{mcond}) with 
Eqs.(\ref{j}) and (\ref{dis}) reflects the following differential equations 
\begin{eqnarray}
&&\xi_1' +\frac{\alpha}{2} \sqrt{-g} =0,\ \ \xi_2' -\frac{\alpha}{2} \sqrt{-g} =0,\ \ \xi_3' -\frac{\alpha}{2} \sqrt{-g} =0,\ \ \xi_4' -\frac{\alpha}{2} \sqrt{-g} =0,\nonumber\\
&& \xi_5' -\frac{\alpha}{2} \sqrt{-g} =0,\ \ \xi_6' +\alpha \sqrt{-g} =0,\ \ \xi_7' + \alpha \sqrt{-g} =0, \label{diffe}
\end{eqnarray}
satisfying the relations
\begin{eqnarray}
&&\xi_1+\xi_2=0,\ \ \xi_1 +\xi_3 =0,\ \ \xi_1 +\xi_4 =0,\ \ \xi_2 -\xi_3=0,\ \ \xi_2 -\xi_4 =0,\nonumber\\
&& \xi_3-\xi_4 =0,\ \ 2\xi_5 +\xi_6 =0,\ \ 2\xi_5 +\xi_7 =0, \ \ \xi_6 -\xi_7 =0.\label{con}
\end{eqnarray}
The solutions of the differential equations given in Eq. (\ref{diffe}), fulfilling the
boundary conditions $\xi_i (\kappa =0) =0$, are
\begin{eqnarray}
&&\xi_1  = -\frac{\alpha}{2} \sqrt{-g}\kappa,\ \ \xi_2 = \frac{\alpha}{2} \sqrt{-g} \kappa,\ \ \xi_3 = \frac{\alpha}{2} \sqrt{-g}\kappa,\ \ \xi_4
= \frac{\alpha}{2} \sqrt{-g} \kappa,\nonumber\\
&& \xi_5 = \frac{\alpha}{2} \sqrt{-g} \kappa,\ \ \xi_6  =-\alpha \sqrt{-g} \kappa,\ \ \xi_7 =- \alpha \sqrt{-g} \kappa. 
\end{eqnarray}
These identifications of $\xi_i (\kappa)$ also satisfy the conditions in Eq. (\ref{con}).
Therefore, the FFBRST transformation (\ref{ff}) with parameter  $\Theta$ obtainable from Eq. (\ref{theta}) changes 
the
effective action within functional integration as
\begin{eqnarray}
S_{NL} +S_1 (\kappa =1) &=&\int d^4 x \left[{\cal L}_c +i \sqrt{-g}b^a(\nabla^b h_{ab}-k \nabla_a h)-i \sqrt{-g}\bar c^b \nabla_b c^a (\nabla^b h_{ab}-k \nabla_a h)\right.\nonumber\\
&+&\left. \sqrt{-g}\bar c^a M_{ab} c^b \right].
\end{eqnarray}
After shifting the Nakanishi--Lautrup field by $\bar c^b \nabla_b c^a $, the above expression reduces to
\begin{eqnarray}
S_{NL} +S_1 (\kappa =1) &=&\int d^4 x \left[{\cal L}_c +i \sqrt{-g}b^a(\nabla^b h_{ab}-k \nabla_a h) + \sqrt{-g}\bar c^a M_{ab} c^b \right],
\nonumber\\
&=&S_L,
\end{eqnarray}
which is nothing but the effective action for perturbative quantum gravity in Landau gauge. 

We end   this section by making the following conclusion that the finite field-dependent BRST with appropriate choice of 
finite parameter connects two different gauges in the theory of quantum gravity in curved spacetime. 
We show these results also at quantum level using BV formulation in the next section.
\section{BV formulation and FFBRST symmetry} 
In  the BV formulation the generating functional   of quantum gravity (in Landau gauge) in curved spacetime,  by  
introducing antifields $\phi^\star $ corresponding to the all fields $\phi ( \equiv h, \bar c, c, b)$
 with opposite statistics, is given by
{\begin{eqnarray}
Z_L  = \int {\cal D}\phi\ e^{ i\int d^4x ( {\cal L}_c +{\cal L}_g [\phi, \phi^\star])}.
\end{eqnarray}}
This can further be written in compact form as
 \begin{equation}
Z_L = \int {\cal D}\phi\  e^{ i    W^L_{\Psi  }[\phi,\phi^\star] },
\end{equation} 
where $ W^L_{\Psi  }[\phi,\phi^\star]$ is an extended quantum action in Landau gauge.
The generating functional does not depend on the choice of gauge-fixing fermion \cite{ht}.
The extended quantum action for perturbative quantum gravity,
 $W_{\Psi }[\phi,\phi^\star]$, satisfies  the following   mathematically rich 
 relation, called the quantum master equation \cite{wei},  
\begin{equation}
\Delta e^{iW_{\Psi }[\phi, \phi^\star ]} =0  \ \mbox{ with }\ 
 \Delta\equiv \frac{\partial_r}{
\partial\phi}\frac{\partial_r}{\partial\phi^\star } (-1)^{\epsilon
+1}.
\label{mq}
\end{equation}
The antifields get identified  with gauge-fixing fermion ($\Psi$),  given in Eq. (\ref{gff}), as
follows
{ \begin{eqnarray}
h_{ab}^{\star }&=&\frac{\delta\Psi }{\delta h^{ab}}= i  \sqrt{- g} (-\nabla_b \bar c_a +kg_{ab}\nabla^c \bar c_c),
\nonumber\\
 \bar c_a^{ \star}&=&\frac{\delta\Psi }{\delta \bar c^a}= i  \sqrt{- g} (\nabla^b h_{ab} -k\nabla_a h),\nonumber\\ c_a^{\star}&=&\frac{\delta\Psi}{\delta c^a}= 
 0,\ \ b_a^{\star}=\frac{\delta\Psi}{\delta b^a}= 
 0.
\end{eqnarray}}
Similarly, the generating functional for quantum gravity in a non-linear gauge  is defined, compactly, as  
{\begin{eqnarray}
Z_{NL}  &=& \int {\cal D}\phi\ e^{ i\int d^4x ( {\cal L}_c +{\cal L'}_g [\phi, \phi^\star])},\nonumber\\
&=& \int {\cal D}\phi\  e^{ i    W^{NL}_{\Psi  }[\phi,\phi^\star] }.
\end{eqnarray}}

The following expression for antifields in the case of a non-linear gauge are obtained
{ \begin{eqnarray}
h_{ab}^{\star } &= &i  \sqrt{- g} (-\nabla_b \bar c_a +kg_{ab}\nabla^c \bar c_c),\nonumber\\
 \bar c_a^{ \star} &=& i  \sqrt{- g}\left(\nabla^b h_{ab} -k\nabla_a h-i\frac{\alpha}{2}\nabla_a \bar c^b c_b 
+i\frac{\alpha}{2} b_a +i\frac{\alpha}{2}\bar c^b \nabla_b c_a\right),\nonumber\\
 b_a^{\star} &= &  i  \frac{\alpha}{2}\sqrt{- g}\bar c_a,\ \ c_a^{\star} =i \frac{\alpha}{2} \sqrt{- g}\left[
 \bar c^b \nabla_b \bar c_a +\nabla_b (\bar c_a \bar c^b)\right].
\end{eqnarray}} 
To connect linear and non-linear gauges in BV formulation we construct the following finite field-dependent parameter $\Theta [\phi]$ 
\begin{eqnarray}
\Theta [\phi] =-\int_0^1 d\kappa \int d^4y \left[c_a^\star c^a -b_a^\star b^a \right].
\end{eqnarray} 
The Jacobian of the path integral measure  in the generating functional for   this FFBRST parameter can be replaced by $e^{iS_1 [\phi, \phi^\star]}$ iff condition (\ref
{mcond}) is satisfied. 
 The factor $e^{iS_1 [\phi, \phi^\star]}$ changes the quantum action as
\begin{eqnarray}
W^{NL}_{\Psi  }[\phi,\phi^\star]    \stackrel{FFBRST}{----\longrightarrow }
W^L_{\Psi  }[\phi,\phi^\star].
\end{eqnarray}
This reflects the validity of results up to quantum levels also. Hence,
we conclude that the finite field-dependent transformations connect two different solutions of quantum master equation also
in the case of quantum gravity  in curved spacetime.

\section{  Conclusions}
In this work we have analyzed the general coordinate invariance of perturbative theory of quantum gravity
with different gauge conditions. 
It has been shown that  the theory of perturbative quantum gravity in general curved 
spacetime dimensions in linear and non-linear gauges has supersymmetric   BRST and anti-BRST invariance.  Further 
we have developed the FFBRST transformation for  quantum gravity  by constructing a general finite and field-dependent parameter. 
 The finite and field-dependent structure of BRST transformation  changes the path integral of quantum gravity
  non-trivially  as in the case of gauge theory in the flat spacetime. 
   We have observed that the results of FFBRST formulation of usual gauge theory also hold 
 in the theory of gravity on curved spacetime. In this context we have shown that the non-trivial Jacobian appearing in the
 functional integration  is responsible for differences in the effective action of perturbative quantum gravity.
 The FFBRST transformation with suitable choice of finite parameter 
 connects the linear Landau and non-linear Curci--Ferrari gauges in the theory of curved spacetime.
 The validity of this result at quantum level is also established by explicit calculations
 in BV formulation. 
  We expect several other application
 of this formulation for the  theory of perturbative quantum gravity. It would be very interesting
 to analyze the ghost and graviton condensation for theory of quantum gravity in the Landau and massive Curci--Ferrari gauges.

\section*{Acknowledgments}
This work is dedicated to the memory of my late brother Ramesh Chandra Upadhyay.


\begin{thebibliography}{0}
\bibitem{hooft} 	G. t Hooft, ``Perturbative Quantum Gravity" From Quarks and Gluons to Quantum Gravity. Vol. 1. (2003).
\bibitem{hata1} T. Hatanaka and S. V. Ketov,  Nucl. Phys. B 794, 495 (2008).
\bibitem{hata2}  T. Hatanaka and S. V. Ketov, Class. Quant. Grav. 23, L45 (2006).
\bibitem{asch}  P. Aschieri, M. Dimitrijevic, F. Meyer and J. Wess, Class. Quant. 
Grav. 23, 1883 (2006).
\bibitem{tsa}  N. C. Tsamis and R. P. Woodard, Phys. Lett.  B 292,  269 (1992). 
\bibitem{fi} M. Faizal,  arXiv:1105.3112.
\bibitem{wood} J. Iliopoulos, T. N. Tomaras, N. C. Tsamis and R. P. Woodard, Nucl. Phys. B 534,  419 (1998).
\bibitem{ein} A. Einstein, The Meaning of Relativity, fifth edition (Princeton University Press, 1956).
\bibitem{ch} A. H. Chamseddine, Int. J. Mod. Phys. A 16, 759 (2001).
\bibitem{da}  T. Damour, S. Deser and J. McCarthy, Phys. Rev. D 47, 1541 (1993).
\bibitem{ah} A. H. Chamseddine, Commun. Math. Phys. 218, 283 (2001).
\bibitem{dud1}D. Dudal, H. Verschelde, V. E. R. Lemes, M. S. Sarandy, S. P. Sorella and M. Picariello, 
Ann. Phys. 308, 62 (2003).
\bibitem{dud2}D. Dudal, V. E. R. Lemes, M. Picariello, M. S. Sarandy, S. P. Sorella and H. Verschelde, 
JHEP. 0212, 008 (2002).  
\bibitem{brst} C. Becchi, A. Rouet and R. Stora,  Annals Phys. {98}, 287 
(1974). 
\bibitem {tyu} I. V. Tyutin, Lebedev Physics Institute preprint 39 (1975), arXiv: 0812.0580.
\bibitem{ht} M. Henneaux and C. Teitelboim,  Quantization of gauge
systems  (Princeton, USA: Univ. Press, 1992).
\bibitem{wei} S. Weinberg,   The quantum theory of fields, Vol-II: Modern
applications (Cambridge, UK Univ. Press, 1996).
\bibitem{bv} I. A. Batalin and G. A. Vilkovisky,  Phys. Lett.  B 102, 
27 (1981).
\bibitem{bv1} I. A. Batalin and G. A. Vilkovisky,  Phys. Rev. D 28, 
2567 (1983); 
Erratum ibid  D 30, 508 (1984).
\bibitem{bv2} I. A. Batalin and G. A. Vilkovisky,  Phys. Lett.  B 120  
166 (1983).
\bibitem{subm}  S. Upadhyay and B. P. Mandal, Eur. Phys. J. C 72, 2059 (2012).
\bibitem{bss} B. P. Mandal, S. K. Rai, and S. Upadhyay, EPL 
92, 21001 (2010).
\bibitem{na} N. Nakanishi, Prog. Theor. Phys. 59, 972 (1978).
\bibitem{ku} T. Kugo and I. Ojima, Nucl. Phys. B 144, 234 (1978).
\bibitem {ni} K. Nishijima and M. Okawa, Prog. Theor. Phys. 60, 272 (1978).
\bibitem{nn} N. Nakanishi and I. Ojima, Covariant operator formalism of gauge theories
and quantum gravity  (World Sci. Lect. Notes. Phys.  1990).
\bibitem{yo} Y. Kitazawa, R. Kuriki and K. Shigura, Mod. Phys. Lett.
A 12, 1871 (1997).
\bibitem{be} E. Benedict, R. Jackiw and H. J. Lee, Phys. Rev. D 54, 6213 (1996).
\bibitem{fr} F. Brandt, W. Troost and A. V. Proeyen, Nucl. Phys.
B464, 353 (1996).
\bibitem{faiza} M. Faizal, Found. Phys. 41, 270  (2011).
\bibitem{upa} S. Upadhyay,  Phys. Lett. B 723, 470 (2013). 
  \bibitem{sdj} S. D. Joglekar and B. P. Mandal,   {Phys. Rev.}  {D 51}, 1919 (1995).
\bibitem{sdj1}  S. D. Joglekar and B. P. Mandal, Int. J. Mod. Phys. A 17, 1279 (2002).
\bibitem{rb} R. Banerjee and B. P. Mandal, Phys. Lett. B 488, 27 (2000).
 \bibitem{susk}   S. Upadhyay,   S. K. Rai and B. P. Mandal,  J. Math. Phys.  {52}, {022301} (2011).
\bibitem{sb} S. Upadhyay and B. P. Mandal,  Eur. Phys. J.  {C 72},  2065 
(2012); Annals of Physics {327}, 2885 (2012); Europhys. Lett.  {93}, 
{31001} (2011); Mod. Phys. Lett.   {A 25}, {3347} (2010). 
\bibitem{smm} S. Upadhyay, M. K. Dwivedi and B. P. Mandal, Int. J. Mod. Phys. A 28, 1350033 (2013).
\bibitem{fs} M. Faizal, B. P. Mandal and S. Upadhyay, Phys. Lett. B 721, 159 (2013).
\bibitem{sud1}  R. Banerjee, B. Paul and S. Upadhyay,  Phys. Rev. D 88, 065
019 (2013).
\bibitem{sudhak} S. Upadhyay,  arXiv:1310.2013 [hep-th],  to appear in Phys. Lett. B (2013).
\bibitem{rbs} R. Banerjee, B. Paul, S. Upadhyay, Phys. Rev. D 88 (2013)  065019.
\bibitem{rs} R. Banerjee and S. Upadhyay,  arXiv:1310.1168 [hep-th].
\bibitem{hig} A. Higuchi and S. S. Kouris, Class. Quant. Grav. 18, 4317 (2001).
 
\end{thebibliography}
\end{document}